\documentstyle[prd,aps,epsf]{revtex}

\def\ucslash#1{/\!\!\!\!#1}

\begin{document}
\draft
\title{Neutrino-electron processes in a strongly magnetized thermal plasma}
\author{Stephen J. Hardy$^{1,2}$ and Markus H. Thoma$^{3,4}$}
\address{$^1$Max-Plank-Institut f\"ur Astrophysik,
Karl-Schwarzschild-Str. 1, 85740 Garching bei M\"unchen,
GERMANY\\$^2$Research Centre for Theoretical Astrophysics, School of
Physics, University of Sydney, 2006 AUSTRALIA}
\address{$^3$Theory Division, CERN, CH-1211 Geneva 23, SWITZERLAND\\
$^4$ Institut f\"ur Theoretische Physik, Universit\"at Giessen, 35392 Giessen,
GERMANY}

\date{Draft version::\today}
\maketitle
\begin{abstract}
We present a new method of calculating the rate of neutrino-electron
interactions in a strong magnetic field based on finite temperature
field theory. Using this method, in which the effect of the magnetic
field on the electron states is taken into account exactly, we
calculate the rates of all of the lowest order neutrino-electron
interactions in a plasma. 
As an example of the use of this technique, we explicitly calculate
the rate at which neutrinos and antineutrinos annihilate in a highly
magnetized plasma, and compare that to the rate in an unmagnetized
plasma. The most important channel for energy deposition is the
gyromagnetic absorption of a neutrino-antineutrino pair on an electron
or positron in the plasma ($\nu\bar{\nu} e^\pm\leftrightarrow
e^\pm$). Our results show that the rate of annihilation increases with
the magnetic field strength once it reaches a certain critical value,
which is dependent on the incident neutrino energies and the ambient
temperature of the plasma. It is also shown that the annihilation
rates are strongly dependent on the angle between the incident
particles and the direction of the magnetic field. If sufficiently
strong fields exist in the regions surrounding the core of a type II
supernovae or in the central engines of gamma ray bursts, these
processes will lead to more efficient plasma heating mechanism than in
an unmagnetized medium, and moreover, one which is intrinsically
anisotropic.
\end{abstract}


\section{Introduction}
Neutrino heating and cooling plays an important role in a variety
of astrophysical objects. In core collapse supernovae (SNe) neutrinos
produced deep within the core of the forming protoneutron star (PNS)
are thought to deposit some fraction of their energy in a
semitransparent region above the surface of the PNS, leading to a
neutrino driven wind \cite{duncan} and a robust supernova explosion
\cite{colgate}. Although it is currently thought that neutrino-nucleon
scattering provides the bulk of energy transfer in this region, a
significant fraction of the energy and momentum exchange between the
neutrinos and the matter occurs through neutrino-antineutrino
annihilation to electron-positron pairs and through neutrino-electron
scattering.

More recently, neutrino-electron interactions have been proposed as an
energy deposition mechanism for the central engines of gamma-ray
bursts (GRBs). A large fraction of the binding energy released during
the formation of a compact object is emitted in the form of neutrinos
and antineutrinos. Both neutron star-neutron star mergers and the new
``collapsar'' models for the formation of GRB fireballs rely on
neutrino-antineutrino annihilation to electron-positron pairs as a
mechanism for the transport of energy from the dense and hot regions
to regions of low baryon loading where a fireball can form and
expand \cite{popham}. Unfortunately, even for the extremely high neutrino densities
expected in these systems, the cross section of neutrino-antineutrino
annihilation in the absence of a strong magnetic field is still quite
low, which means that the overall efficiency of conversion of
gravitational potential energy to fireball energy is also low. Thus,
it is very difficult to explain the most energetic of GRBs using 
neutron star-neutron star
merger models, and even the collapsar models have some difficulty in
depositing sufficient energy to drive the fireball through the
overlying mantle of the star \cite{janka}.

The astrophysical arguments for the existence of supercritical fields
in nature have grown stronger recently with the discovery of
magnetars \cite{hurley}. In these slowly rotating X-ray emitters, field strengths of
up to $4 \times 10^{15} \,\mbox{G}$ have been inferred. One may
speculate that even stronger fields may be present when such objects
are born. This discovery negates previous theoretical
prejudice against the existence of such strong fields, and suggest
further examination of the effect that strong magnetic fields may have
on neutrino-electron processes.

In the absence of a strong magnetic field, there are only two allowed
types of interaction between neutrinos and electrons --
neutrino-electron scattering ($\nu e^\pm \leftrightarrow \nu e^\pm$,
$\bar{\nu} e^\pm \leftrightarrow \bar{\nu} e^\pm$), and
electron-positron pair creation and annihilation
($\nu\bar{\nu}\leftrightarrow e^+ e^-$). In the presence of a
quantizing magnetic field, electrons occupy definite states of
momentum perpendicular to the field (Landau levels) and conservation
of perpendicular momentum between interacting particles is no longer
required, as some of the momentum may be absorbed by the field. This
allows a number of exotic reactions to proceed, where electrons or
positrons jump between different Landau levels while interacting with
an external neutrino current. These processes are very similar to
interactions between photons and electrons in a strong field, such as
single photon pair creation, which is also forbidden in the absence
of a strong field. The additional neutrino processes allowed in the
presence of a strong magnetic field are absorption of a
neutrino-antineutrino pair by an electron or positron ($\nu\bar{\nu}
e^\pm\leftrightarrow e^\pm$), and the absorption or emission of an
electron positron pair by a neutrino or antineutrino ($\nu e^+ e^-
\leftrightarrow \nu $, $\bar{\nu} e^+ e^- \leftrightarrow
\bar{\nu}$). Each of these may have an important effect on the
neutrino energy exchange opacities in a strongly magnetized plasma.

A number of authors have shown that the rate of neutrino-electron
processes are increased due to the presence of a supercritical
magnetic field ($B > B_{cr} = 4.4 \times 10^{13}
\,\mbox{G}$). Kuznetsov and Mikheev \cite{kuznetsov} considered
neutrino-electron scattering and the emission of an electron positron
pair by a neutrino propagating through a magnetic field in the limit
that the electrons and positrons are in the lowest Landau level. They
concluded that with a very strong field, the neutrino could lose only a
small fraction of its energy through this process. While we do not
analyse this process in detail numerically, our results for the
neutrino-antineutrino annihilation processes suggest that the
restriction that the electron and positron be in the lowest Landau
level is too strong. A more general calculation may be performed
using the results presented in this paper.  Benesh and Horowitz
\cite{benesh} calculated the rate of neutrino-antineutrino
annihilation to electron-positron pairs in a strong magnetic
field. They examined two restrictive cases - nearly parallel
collisions and head-on collisions - and found no significant
enhancement of the cross section for these types of interactions. Our
more general calculation shows that there are regions of phase space
with a significant increase in the annihilation cross-section, though
for moderate field strengths, these regions are small. More
importantly, we show that the new channels for annihilation which
appears in a strong magnetic field - gyromagnetic absorption on
electrons and positrons in the plasma - are more important over an
astrophysically interesting range of phase space.

Bezchastnov and Haensel \cite{bezch} calculated the rate of
neutrino-electron scattering in a strong magnetic field by evaluating
the matrix element for the Feynman diagrams shown in figure 1, using
the exact wave-functions of the electrons in the strong field as
external states for the calculation. Considering energy ranges and
magnetic field strengths appropriate for SNe calculations, they showed
that while the overall cross-section for this scattering rate was not
a strong function of magnetic field strength up to $10^{15}$G, the
strong field introduced significant anisotropies in the rate, and could
lead to interesting parity violating effects.

Here we adopt a new approach to calculating the rates of
neutrino-electron processes in a strong magnetic field.  Rather than
adopting the direct approach of calculating the cross-sections from
the diagrams shown in figure 1, such as adopted in \cite{benesh} and
\cite{bezch}, we use a technique of finite temperature field theory
(FTFT) which relates the interaction rates to the imaginary part of
the diagram shown in figure 2 (as the energy of our interactions are
low compared to the $W$-boson mass, we use the Fermi theory to
describe our interactions). This technique allows all the
neutrino-electron interaction rates to be calculated on the same
footing, and includes the effects of Pauli blocking. We follow the
approach of Gale and Kapusta \cite{gale} which was developed in the
context of calculating the rate of dilepton production from hadron
interactions in heavy ion collisions. It is based on the fact that the rate
of dilepton production is related to the imaginary part of the
polarization tensor. In general, production as well as decay rates of
particles in matter can be extracted from the imaginary part of self
energies computed by using FTFT \cite{weldon1}.  In our case, the
interacting particles are neutrinos, not hadrons, and our interactions
are electroweak interactions and not due to nuclear forces. This
approach leads to two important simplifications, in that one obtains
the total net rate of a given process, taking into account both
forward and back reactions. Also, numerical evaluation of the rates is
greatly simplified, as one can calculate all of the rates virtually
simultaneously.

The plan of this paper is as follows. In the first section, the
expression relating the rate of a given neutrino-electron process to
the imaginary part of the polarization tensor, based on the work of
Weldon \cite{weldon1} and Gale and Kapusta \cite{gale}, is presented.
This expression is then used to calculate the rate at which neutrinos
and antineutrinos annihilate to electron-positron pairs in an
unmagnetized vacuum, demonstrating that one obtains the same results
as the direct calculation.

In the following section, a strong magnetic field is introduced, and
the polarization tensor in a strongly magnetized plasma is
presented. The imaginary part of the polarization tensor is
calculated, and expressions for the rates of each of the
neutrino-electron processes are given. A detailed numerical evaluation
of the annihilation rate is then performed, and the rate is shown to
reproduce the unmagnetized rate in the limit of a weak magnetic field.

Finally, the behaviour of the annihilation rate with increasing
magnetic field strength is examined, and the implications for
systems in which significant neutrino heating occurs, such as
core collapse SNe and GRBs are discussed.

\begin{figure}
\centerline{\epsfxsize=5cm\epsfbox{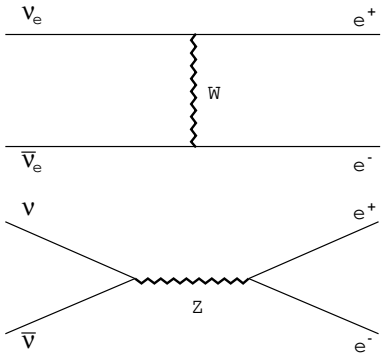}}
\caption{Feynman diagram for electron-positron pair creation through
neutrino-antineutrino annihilation.}
\end{figure}

\begin{figure}
\centerline{\epsfxsize=5cm\epsfbox{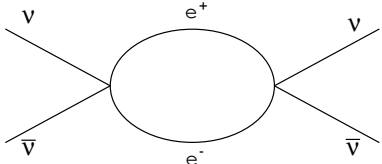}}
\caption{This is the diagrammatic representation of the polarization
tensor used to calculate the rate of electron-positron pair production
through neutrino-antineutrino annihilation. The imaginary part of this
diagram is related to the rate of the process shown in figure 1.
}
\end{figure}

\section{Basic formalism}

Ultimately, we wish to calculate the rate at which neutrinos and
antineutrinos interact with electrons and positrons in a strong
magnetic field in matter through the processes shown in figure 1. 
Rather than
calculate these rates directly through the use of the exact electron wave
functions in the magnetic field, such as performed by Bezchastnov
and Haensel \cite{bezch} in their evaluation of the neutrino-electron
scattering rate, we adopt an alternative approach suggested by the use
of finite temperature field theory (FTFT).
This theory has been pioneered by Weldon
\cite{weldon1} and also applied by Gale and Kapusta \cite{gale} to calculate
the rate at which electron-positron pairs are emitted from the fireball
in heavy ion collisions. 

According to Weldon \cite{weldon1} the decay and production rate of 
a boson with energy $E$ at temperature $T$ can be related to the 
imaginary part of the boson self energy $\Pi $ by
\begin{equation} 
\Gamma (E)=-\frac{1}{E}\> {\rm Im}\{\Pi (E)\},
\label{gamma}
\end{equation}
where $\Gamma =\Gamma_d-\Gamma_i$ is the difference of the forward
rate of the boson decay $\Gamma_d$ and the rate of the inverse
process $\Gamma_i$.  The decay and production rates are related by the
principle of detailed balance
\begin{equation}
\frac{\Gamma_d}{\Gamma_i}=e^{E/T},
\label{detbal}
\end{equation}
from which we get
\begin{equation} 
\Gamma_d(E)=-\frac{1}{E}\frac{1}{1-e^{-E/T}}\> {\rm Im} \{\Pi (E)\}.
\label{gamma2}
\end{equation}

Instead of considering the decay of a boson we will investigate first 
the case of neutrino-antineutrino annihilation to an
electron-positron pair, $\nu \bar{\nu} \rightarrow
e^- e^+$, in an electron-positron plasma at temperature $T_e$ and chemical 
potential $\mu_e$. The corresponding annihilation 
rate is given by equation (\ref{gamma2}),
where the boson is replaced by the neutrino-antineutrino pair.
To lowest order the self energy is now given by the diagram of figure
2 which contains an electron loop. The four momenta of the neutrino
and antineutrino are denoted by $Q_1$ and $Q_2$ and of the
electron and positron by $P_1$ and $P_2$, respectively. Here we use the 
notation $Q_i=(q_i,{\bf q_i})$ with $q_i=|{\bf q_i}|$ and 
$P_i=(E_i,{\bf p_i})$ with $E_i=\sqrt{p_i^2+m^2}$, where $m$ is
the electron mass. Then in equation (\ref{gamma2}) $E=q_1+q_2$ holds and 
the self energy is given by a Lorentz tensor $\Pi _{\mu \nu}$.

Next we want to relate the annihilation rate to the differential rate 
$dR_{\nu \bar \nu}/{d^3q_1d^3q_2}$ describing the absorption of 
a neutrino-antineutrino pair with momenta ${\bf q_1}$ and ${\bf q_2}$
in the electron-positron plasma. Usually, this rate would be calculated from the matrix element
${\cal M}$ of figure 1 according to
\begin{eqnarray}
\frac{dR_{\nu \bar \nu}}{d^3q_1d^3q_2} &&= \frac{1}{2q_12q_2} \int
\frac{d^3p_1}{2E_1(2\pi )^3}[1-n_{e^+}(E_1)] \int \frac{d^3p_2}{2E_2(2\pi
)^3}[1-n_{e^-}(E_2)] \nonumber \\ && (2\pi)^4\delta^4(Q_1+Q_2-P_1-P_2)
\sum_{i} \langle |{\cal M}|^2 \rangle,
\label{matrix}
\end{eqnarray}
where there is a sum over initial states and the Fermi distribution 
function for the electron is given by
\begin{equation}
n_{e^-}(E)=\frac{1}{e^{(E-\mu_e)/T_e}+1}
\label{fermi}
\end{equation}
and the one of the positron by replacing $\mu_e$ by $-\mu_e$ in
equation (\ref{fermi}). The distribution functions in equation
(\ref{matrix}) describe Pauli blocking of the produced electron and
positron in the plasma.

Following Gale and Kapusta \cite{gale} the differential rate is obtained
from the annihilation rate $\Gamma_d$ by taking into account the
neutrino current 
\begin{equation}
\frac{dR_{\nu \bar \nu}}{d^3q_1d^3q_2} 
=\frac{1}{2q_12q_2} \frac{1}{1-e^{-E/T_e}} M^{\mu \nu}\>  {\rm Im} 
\{\Pi_{\mu \nu}(Q_1+Q_2)\}.
\label{diffrate}
\end{equation}
Here the leptonic tensor for the neutrinos is given by
\begin{eqnarray}
M^{\mu\nu} & = & \sum _{s_1s_2} \bar{v}(q_2) \gamma^\mu(1-\gamma_5)
u(q_1) \bar{u}(q_1) \gamma^\nu(1-\gamma_5)v(q_2) \nonumber \\
& = & 8(Q_2^\mu Q_1^\nu + Q_1^\mu Q_2^\nu - (Q_1 Q_2) g^{\mu\nu} + i
Q_{1\alpha} Q_{2\beta} \varepsilon^{\alpha\mu\beta\nu}),
\label{leptens} 
\end{eqnarray}
where we have summed over the neutrino spins. This formula differs from 
the one for the dilepton production rate
given in the appendix of ref. \cite{gale} by the absence of  
a photon propagator and a different structure of the leptonic tensor
for massless neutrinos instead of electrons leading to the
appearance of the antisymmetric tensor $\varepsilon^{\alpha\mu\beta\nu}$
in equation (\ref{leptens}). Furthermore the two expressions differ by a factor
$\exp{(E/T_e)}$ according to (\ref{detbal})
because we are considering a decay instead of a production rate. 

The total rate for the production of electron-positron pairs from 
neutrino-antineutrino annihilation follows by integrating the
differential forward rate, equation (\ref{diffrate}), 
over the neutrino distribution functions $f_{\nu,\bar \nu}(q_i)$
\begin{equation}
R_{\nu \bar \nu}=\int \frac{d^3q_1}{(2\pi )^3}\int \frac{d^3q_2}{(2\pi )^3}
[f_\nu(q_1)f_{\bar \nu}(q_2) - (1-f_\nu(q_1))(1-f_{\bar \nu}(q_2)) e^{-E/T_e}]\>  
\frac{dR_{\nu \bar \nu}}{d^3q_1d^3q_2}.
\label{totrate}
\end{equation}
Here we have included the back reaction $e^-e^+ \rightarrow \nu \bar \nu$,
which is described by the second term in the square brackets.
It should be noted that our derivation of the rates at no time
assumed that the neutrinos are in thermal equilibrium with the
electron-positron plasma. For example, neutrinos that escape from the
core of a supernova explosion may have a roughly thermal distribution with a temperature
of $T_\nu \simeq 5$ MeV. After leaving the core these neutrinos will
interact with the surrounding electron-positron plasma which has a 
typical temperature of about $T_e\simeq 1$ MeV. In this case the neutrino
distribution functions in equation (\ref{totrate}) are given by the equilibrium
Fermi distributions, equation (\ref{fermi}), where the temperature and the 
chemical potential are replaced by the corresponding values for the neutrinos
escaping from the core, $T_\nu$ and $\mu_\nu$.

One may also determine the rate at which a neutrino scatters in a
thermal plasma in a similar manner. The differential rate at which 
a neutrino (or antineutrino) scatters from an initial state $Q_1$ to a 
final state $Q_2$, is given simply by replacing $Q_1+Q_2$ by $Q_1-Q_2$
in ${\rm Im} \{\Pi_{\mu \nu}\}$ and $E=q_1+q_2$ by $q_1-q_2$ in 
equation (\ref{diffrate}). The total rate of scattering,$R_{\nu e^\pm}$, is then given
by
\begin{equation}
R_{\nu e^\pm}=\int \frac{d^3q_1}{(2\pi )^3}\int \frac{d^3q_2}{(2\pi )^3}
[f_\nu(q_1)(1-f_\nu(q_2)) - f_\nu(q_2)(1-f_\nu(q_1)) e^{-E/T_e}]\>  
\frac{dR_{\nu e^\pm}}{d^3q_1d^3q_2}
\label{scattotrate}
\end{equation}
with
\begin{equation}
\frac{dR_{\nu e^\pm}}{d^3q_1d^3q_2} 
=\frac{1}{2q_12q_2} \frac{1}{1-e^{-E/T_e}} M^{\mu \nu}\>  {\rm Im} 
\{\Pi_{\mu \nu}(Q_1-Q_2)\},
\label{scatdiffrate}
\end{equation}
where $E = q_1 - q_2$. Note that both the electron and positron
scattering rates are included within the imaginary part of the
polarisation tensor in equation (\ref{scatdiffrate}). The strength
of this method lies in the fact that the correct calculation of the
imaginary part of the polarisation tensor leads the differential
annihilation and scattering rates for all of the separate channels
that are allowed within the plasma. Thus, if one includes the effect
of the magnetic field in the polarisation tensor properly, all of the
exotic processes which occur in a magnetic field are included in the
rate calculations by default.

\subsection{The unmagnetized rate}
To demonstrate that the two approaches, based either on the matrix
element as in equation (\ref{matrix}) 
or on the FTFT as in equation (\ref{diffrate}) are 
equivalent, we use equation (\ref{diffrate}) to calculate the 
rate of neutrino-antineutrino
annihilation to electron positron pairs in an unmagnetized vacuum at
zero temperature, and
show that it reproduces the results of the calculation of the
process using the matrix element method.

In the one-loop approximation
the polarisation tensor in equation (\ref{diffrate}) is given by
\begin{equation}
\Pi^{\mu\nu}(K) = - {G_F^2 \over 2} {\rm Tr} \int {d^4 P \over (2 \pi)^4}
\gamma^\mu (c_{V} - c_{A} \gamma_5) G(P) \gamma^\nu (c_{V} - c_{A} \gamma_5) G(P-K),\label{eq:3}
\end{equation}
where $G_F$ is the Fermi constant, $G(P)$ is an electron propagator,
and
\begin{equation}
c_{V} = \pm {1\over 2} + 2 \sin^2\theta_W, \qquad c_{A} = \pm
{1\over 2} \label{eq:4}
\end{equation}
with $\theta_W$ the Weinberg angle, and the plus sign corresponding to
electron neutrinos, and the minus sign to muon and tau neutrinos.
This tensor has been formulated in the limit that the neutrino
energies are small compared to the $W$ mass, which will always be
justified in practice. This allows the use of effective Fermi
4-vertices in the processes shown in figure 1 and 2. 
The polarization tensor in equation (\ref{eq:3}) may be decomposed
into vector-vector, axial-vector, and axial-axial parts through
\begin{equation}
\Pi^{\mu\nu} = c_{V}^2 \alpha^{\mu\nu} + 2 c_{V} c_{A}
\alpha^{\mu\nu}_5 + c_{A}^2 \alpha^{\mu\nu}_{55}\label{eq:5}
\end{equation}
with
\begin{equation}
\alpha^{\mu\nu} =  - {G_F^2 \over 2} {\rm Tr} \int {d^4 P \over (2 \pi)^4}
\gamma^\mu  G(P) \gamma^\nu  G(P-K),\label{eq:6}
\end{equation}
\begin{equation}
\alpha^{\mu\nu}_5 =  {G_F^2 \over 2} {\rm Tr} \int {d^4 P \over (2 \pi)^4}
{1 \over 2}\left[\gamma^\mu \gamma_5 G(P) \gamma^\nu  G(P-K) +
\gamma^\mu G(P) \gamma^\nu \gamma_5 G(P-K)\right],
\label{eq:7}
\end{equation}
and
\begin{equation}
\alpha^{\mu\nu}_{55} =  - {G_F^2 \over 2} {\rm Tr} \int {d^4 P \over (2 \pi)^4}
\gamma^\mu \gamma_5 G(P) \gamma^\nu \gamma_5 G(P-K). \label{eq:8}
\end{equation}

In this section we will neglect matter contributions to the rate,
i.e., we set $T_e=0$ in equation (\ref{diffrate}), set the
distribution function to zero and use only the vacuum part of the
polarization tensor, equation (\ref{eq:3}). This corresponds to
neglecting the effects of Pauli blocking on the final state particles
in equation (\ref{matrix}).  Inserting only the vacuum part of the
electron propagators, the vacuum part of the polarization tensor,
equation (\ref{eq:3}), may be written
\begin{equation}
\Pi^{\mu\nu}(K) = - {G_F^2 \over 2} {\rm Tr} \int {d^4 P \over (2
\pi)^4} {\gamma^\mu (c_{V} - c_{A} \gamma_5)(\ucslash{P}+ m)
\gamma^\nu (c_{V} - c_{A} \gamma_5)(\ucslash{P} - \ucslash{K} + m)
\over \left[P^2 - m^2 + i0 \right]\left[(P-K)^2 - m^2 +
i0\right]}.\label{eq:9}
\end{equation}
Note that the real part of this tensor must be renormalized due to the
singularities in the denominator of the integrand, but the imaginary
part is finite.

The imaginary part of $\Pi^{\mu\nu}$ (actually, the anti-hermitian
part) is related to its discontinuity across a branch cut. The branch
cut is related to poles that appear above the pair creation threshold,
$K^2 = 4 m^2$, and it lies along the positive $K^2$ axis from $4 m^2$
to infinity. One has
\begin{equation}
{\rm Im}\{\Pi^{\mu\nu}(K)\} = - {G_F^2 \over 32 \pi^2} I^{\mu\nu}
\end{equation}
with
\begin{eqnarray}
I^{\mu\nu} &=& \int d^4 P \> {\rm Tr} \left[\gamma^\mu (c_{V} - c_{A}
\gamma_5)(\ucslash{P} + m) \gamma^\nu (c_{V} - c_{A} \gamma_5)
(\ucslash{P} - \ucslash{K} + m)\right] \nonumber \\ 
& & \qquad\times \delta(P^2 - m^2) \,
\delta((P-K)^2 - m^2). \label{eq:10}
\end{eqnarray}
Inserting a delta function into equation~(\ref{eq:10}) and relabelling
$P \rightarrow P_1$, leads to
\begin{eqnarray}
I^{\mu\nu} & = & -\int d^4 P_1 \int d^4 P_2 \> {\rm Tr} \left[\gamma^\mu
(c_{V} - c_{A} \gamma_5) (\ucslash{P_1} + m) \gamma^\nu (c_{V} - c_{A}
\gamma_5) (\ucslash{P_2} - m)\right] \nonumber \\ & & \qquad \times \delta(P_1^2 - m^2) \,
\delta(P_2^2 - m^2) \delta^4(P_2 + P_1 - K). \label{eq:11}
\end{eqnarray}

The trace in the integrand of equation~(\ref{eq:11}) may be
recognized as proportional to the electron leptonic tensor,
$L^{\mu\nu}$, and the rate of pair production, equation~(\ref{diffrate}),
is then given by
\begin{equation}
{d R \over d^3q_1d^3q_2} = {G_F^2 \over 2} \int {d^3p_1
\over (2\pi)^3 }\int {d^3p_2 \over (2\pi)^3 }
{L_{\mu\nu}M^{\mu\nu} \over 2 E_1\, 2 E_2 \, 2 q_1\,2 q_2} (2\pi)^4 \,\delta^4(P_1+P_2-Q_1-Q_2),
\label{eq:13}
\end{equation}
where
\begin{equation}
\int d^4 P_i \,\delta(P_i^2 - m^2) = \int {d^3p_i\over E_i}
\end{equation}
has been used.

Equation~(\ref{eq:13}) is identical to the rate, equation (4), 
directly calculated for the
process shown in figure 1 by electroweak theory,
using $L_{\mu \nu}M^{\mu \nu}=\sum_i \langle |{\cal M}|^2 \rangle$. 
Thus we have shown
that the procedure outlined above is identical to the standard
method of calculating rates.

For later reference, the rate at which the annihilation process
proceeds in the absence of a magnetic field at zero temperature 
and in the limit of a vanishing electron mass, $m=0$, is given by
\begin{equation}
{d R \over d^3q_1d^3q_2} = {2 G_F^2 (c_V^2 + c_A^2)
\over 3 \pi} q_1^2 q_2^2 (1 - \cos\Theta)^2,\label{eq:34}
\end{equation}
where $\Theta$ is the angle between the neutrino and antineutrino.

\section{Strong magnetic fields}
We may now make a significant generalization of our theory, which
justifies its introduction. The essential point is that the
polarization tensor used in equation~(\ref{diffrate}) contains all the
information about the electronic part of the process, without
reference to the presence or lack of a medium, or the presence or lack
of a magnetic field. All that is required to calculate the correct
rate is the appropriate polarization tensor for the given plasma. To
determine the rate of neutrino-electron processes in a strongly
magnetized plasma, one must use the polarization tensor for a strongly
magnetized plasma. Thankfully, for our purposes, the polarization
tensors in a strong magnetic field are well known, the vector-vector
part was calculated in the form used here by Melrose and coworkers
\cite{melrose}, the vector-axial part was calculated by Kennett and
Melrose \cite{kennett}, and the axial-axial part may be calculated in a
manner directly analogous to both.

The polarization tensor is calculated in \cite{melrose} and \cite{kennett} using a vertex
formulation of the coupling between an electron in a magnetic field
and an external 4 current. The various parts of the polarization
tensor can be expressed in terms of the vector vertex function
$[\Gamma^{\epsilon'\epsilon}_{q'q}(K)]^\mu$, and the axial vertex
function, $\ _5[\Gamma^{\epsilon'\epsilon}_{q'q}(K)]^\mu$, where $q'$
and $q$ denote the quantum numbers of the particles on either side of
the interaction ($p_\parallel$, momentum, $\sigma$, spin, $n$, Landau
orbital), and $\epsilon$ and $\epsilon'$ denote the nature of the
particles, with a plus sign for electrons (the particles) and a minus
sign for the positrons (the antiparticles). The various components of
the polarization tensor may be written,
\begin{eqnarray}
\alpha^{\mu\nu}(K) & = & - {G_F^2  eB \over 4 \pi} \sum_{n',n=0}^{\infty}
\sum_{\epsilon',\epsilon=\pm1} \nonumber \\
& & \qquad\times\int{d p_\parallel \over 2 \pi} {\left\{{1\over 2}(\epsilon' - \epsilon) +
\epsilon n^\epsilon_q - \epsilon' n^{\epsilon'}_{q'}\right\} \over
\omega - \epsilon \varepsilon_q + \epsilon' \varepsilon_{q'} + i0}
T^{\mu\nu}_{\epsilon'\epsilon}, \label{eq:14}
\end{eqnarray}

\begin{eqnarray}
\alpha^{\mu\nu}_5(K) & = &  {G_F^2  eB \over 4 \pi} \sum_{n',n=0}^{\infty}
\sum_{\epsilon',\epsilon=\pm1} \nonumber \\
& & \qquad\times\int{d p_\parallel \over 2 \pi} {\left\{{1\over 2}(\epsilon' - \epsilon) +
\epsilon n^\epsilon_q - \epsilon' n^{\epsilon'}_{q'}\right\} \over
\omega - \epsilon \varepsilon_q + \epsilon' \varepsilon_{q'} + i0}
\,_5T^{\mu\nu}_{\epsilon'\epsilon},\label{eq:15}
\end{eqnarray}

\begin{eqnarray}
\alpha^{\mu\nu}_{55}(K) & = & - {G_F^2  eB \over 4 \pi} \sum_{n',n=0}^{\infty}
\sum_{\epsilon',\epsilon=\pm1} \nonumber \\
& & \qquad\times\int{d p_\parallel \over 2 \pi} {\left\{{1\over 2}(\epsilon' - \epsilon) +
\epsilon n^\epsilon_q - \epsilon' n^{\epsilon'}_{q'}\right\} \over
\omega - \epsilon \varepsilon_q + \epsilon' \varepsilon_{q'} + i0}
\,_{55}T^{\mu\nu}_{\epsilon'\epsilon}\label{eq:16}
\end{eqnarray}
with
\begin{equation}
T^{\mu\nu}_{\epsilon'\epsilon} = \sum_{\sigma',\sigma=\pm1}
[\Gamma^{\epsilon'\epsilon}_{q'q}({\bf k})]^\mu[\Gamma^{\epsilon'\epsilon}_{q'q}({\bf k})]^{\nu*},\label{eq:17}
\end{equation}
\begin{equation}
\,_5T^{\mu\nu}_{\epsilon'\epsilon} = \sum_{\sigma',\sigma=\pm1}
{1 \over 2} \left\{\,_5[\Gamma^{\epsilon'\epsilon}_{q'q}({\bf
k})]^\mu[\Gamma^{\epsilon'\epsilon}_{q'q}({\bf k})]^{\nu*}
+ [\Gamma^{\epsilon'\epsilon}_{q'q}({\bf k})]^\mu\,_5[\Gamma^{\epsilon'\epsilon}_{q'q}({\bf k})]^{\nu*}\right\},\label{eq:18}
\end{equation}
\begin{equation}
\,_{55}T^{\mu\nu}_{\epsilon'\epsilon} = \sum_{\sigma',\sigma=\pm1}
\,_5[\Gamma^{\epsilon'\epsilon}_{q'q}({\bf k})]^\mu\,_5[\Gamma^{\epsilon'\epsilon}_{q'q}({\bf k})]^{\nu*},\label{eq:19}
\end{equation}
and where $K = (\omega,{\bf k})$, $n^\epsilon_q$ are the
electron distribution functions, and $\varepsilon_q$ denotes the energy of the particle in the
magnetic field,
\begin{equation}
\varepsilon_q \equiv \varepsilon(p_\parallel,n) = \left(m^2 + p_\parallel^2
+ 2 n e B\right)^{1/2}.\label{eq:20}
\end{equation}
Implicit throughout equations (\ref{eq:14}) to (\ref{eq:19}) is the
relation $\epsilon' p_\parallel' = \epsilon p_\parallel -
k_\parallel$. Note also that we are using natural units with all
physical quantities scaled against the electron mass.

The polarization tensors given in equations~(\ref{eq:14}) to~(\ref{eq:16})
contain the contributions from both the vacuum polarization and the
electrons and positrons in the plasma. The vacuum part is given by
the term ${1 \over 2}(\epsilon' - \epsilon)$ in the numerator of the
integrand. The real part of this term is divergent and must be
renormalized. We are concerned here only with
the imaginary part, which is finite. Hence, we perform no
renormalization here. The electrons and positrons of the plasma contribute to the
polarization tensor through the distribution functions (more properly, the
occupation numbers) $n^\epsilon_q$, which in a magnetized thermal
medium are given by the Fermi distributions
\begin{equation}
n^\epsilon_q \equiv n^\epsilon(p_\parallel,n) =
\frac{1}{e^{(\varepsilon(p_\parallel,n)-\epsilon\mu_e)/T_e}+1}.
\end{equation}

The general form of the magnetic vertex functions are given in \cite{kennett}, and
a general discussion of the properties of the vector part may be
found in \cite{melrose}. The components of the vertex functions are
\begin{eqnarray}
[\Gamma^{\epsilon \epsilon'}_{q'q}({\bf k})]^0 & = & C_{q'} C_q
[\delta_{\sigma',\sigma} (\delta_{\epsilon',\epsilon} {\cal B} +
\sigma \delta_{\epsilon',-\epsilon}{\cal A}) (J^{l}_{l'-l} + \rho_{n'} \rho_n J^{l+\sigma}_{l'-l})
\nonumber \\
& & - \epsilon \delta_{\sigma',-\sigma}
(\sigma \delta_{\epsilon',\epsilon} {\cal A} - \delta_{\epsilon',-\epsilon}{\cal B})(-\rho_n J^{l+\sigma}_{l'-l-\sigma} + \rho_{n'}
J^{l}_{l'-l-\sigma})],
\label{eq:21}
\end{eqnarray}
\begin{eqnarray}
[\Gamma^{\epsilon \epsilon'}_{q'q}({\bf k})]^1 & = & C_{q'} C_q
[\delta_{\sigma',\sigma} \epsilon (\delta_{\epsilon',\epsilon} {\cal D} +
\sigma \delta_{\epsilon',-\epsilon}{\cal C})
(-\rho_n J^{l+\sigma}_{l'-l-\sigma} - \rho_{n'} J^{l}_{l'-l+\sigma})
\nonumber \\
& &  - \delta_{\sigma',-\sigma} (\sigma \delta_{\epsilon',\epsilon} {\cal C} -
 \delta_{\epsilon',-\epsilon}{\cal D}) (J^{l}_{l'-l} - \rho_{n'} \rho_n  J^{l+\sigma}_{l'-l-2\sigma})],
\end{eqnarray}
\begin{eqnarray}
[\Gamma^{\epsilon \epsilon'}_{q'q}({\bf k})]^2 & = & i C_{q'} C_q
[\delta_{\sigma',\sigma} \epsilon (\sigma \delta_{\epsilon',\epsilon} {\cal D} +
\delta_{\epsilon',-\epsilon}{\cal C})
(\rho_n J^{l+\sigma}_{l'-l-\sigma} - \rho_{n'} J^{l}_{l'-l+\sigma})
\nonumber \\
& & - \delta_{\sigma',-\sigma} (\delta_{\epsilon',\epsilon} {\cal C} -
 \sigma \delta_{\epsilon',-\epsilon}{\cal D})(J^{l}_{l'-l} + \rho_{n'} \rho_n  J^{l+\sigma}_{l'-l-2\sigma})],
\end{eqnarray}
\begin{eqnarray}
[\Gamma^{\epsilon \epsilon'}_{q'q}({\bf k})]^3 & = & C_{q'} C_q
[\delta_{\sigma',\sigma} (\delta_{\epsilon',\epsilon} {\cal A} +
\sigma \delta_{\epsilon',-\epsilon}{\cal B})
(J^{l}_{l'-l} + \rho_{n'}\rho_n J^{l+\sigma}_{l'-l})
\nonumber \\
& & -\epsilon \delta_{\sigma',-\sigma}(\sigma \delta_{\epsilon',\epsilon} {\cal B} -
\delta_{\epsilon',-\epsilon}{\cal A})
(-\rho_n J^{l+\sigma}_{l'-l-\sigma} + \rho_{n'} J^{l}_{l'-l-\sigma})],
\label{eq:22}
\end{eqnarray}
and
\begin{eqnarray}
\,_5[\Gamma^{\epsilon \epsilon'}_{q'q}({\bf k})]^0 & = & C_{q'} C_q
[\delta_{\sigma',\sigma} (\sigma \delta_{\epsilon',\epsilon} {\cal A} +
\delta_{\epsilon',-\epsilon}{\cal B})
(J^{l}_{l'-l} - \rho_{n'} \rho_n J^{l+\sigma}_{l'-l})
\nonumber \\
& & - \epsilon \delta_{\sigma',-\sigma} (-\delta_{\epsilon',\epsilon} {\cal B} +
\sigma \delta_{\epsilon',-\epsilon}{\cal A})
(-\rho_n J^{l+\sigma}_{l'-l-\sigma} - \rho_{n'} J^{l}_{l'-l-\sigma})],
\label{eq:23}\end{eqnarray}
\begin{eqnarray}
\,_5[\Gamma^{\epsilon \epsilon'}_{q'q}({\bf k})]^1 & = & C_{q'} C_q
[\delta_{\sigma',\sigma} \epsilon( \sigma \delta_{\epsilon',\epsilon} {\cal C} +
\delta_{\epsilon',-\epsilon}{\cal D})
(-\rho_n J^{l+\sigma}_{l'-l-\sigma} + \rho_{n'} J^{l}_{l'-l+\sigma})
\nonumber \\
& & - \delta_{\sigma',-\sigma} (-\delta_{\epsilon',\epsilon} {\cal D} +
\sigma \delta_{\epsilon',-\epsilon}{\cal C})
(J^{l}_{l'-l} + \rho_{n'} \rho_n  J^{l+\sigma}_{l'-l-2\sigma})],
\end{eqnarray}
\begin{eqnarray}
\,_5[\Gamma^{\epsilon \epsilon'}_{q'q}({\bf k})]^2 & = & i C_{q'} C_q
[\delta_{\sigma',\sigma} \epsilon ( \delta_{\epsilon',\epsilon} {\cal C} +
\sigma \delta_{\epsilon',-\epsilon}{\cal D})
(\rho_n J^{l+\sigma}_{l'-l-\sigma} + \rho_{n'} J^{l}_{l'-l+\sigma})
\nonumber \\
& & - \delta_{\sigma',-\sigma} ( - \sigma \delta_{\epsilon',\epsilon} {\cal D} +
\delta_{\epsilon',-\epsilon}{\cal C})
(J^{l}_{l'-l} - \rho_{n'} \rho_n  J^{l+\sigma}_{l'-l-2\sigma})],
\end{eqnarray}
\begin{eqnarray}
\,_5[\Gamma^{\epsilon \epsilon'}_{q'q}({\bf k})]^3  &=&  C_{q'} C_q
[\delta_{\sigma',\sigma} (\sigma \delta_{\epsilon',\epsilon} {\cal B} +
\delta_{\epsilon',-\epsilon}{\cal A})
(J^{l}_{l'-l} - \rho_{n'}\rho_n J^{l+\sigma}_{l'-l})
\nonumber \\
& & - \delta_{\sigma',-\sigma} \epsilon (-\delta_{\epsilon',\epsilon} {\cal A} +
\sigma \delta_{\epsilon',-\epsilon}{\cal B})
(-\rho_n J^{l+\sigma}_{l'-l-\sigma} - \rho_{n'} J^{l}_{l'-l-\sigma})],
\label{eq:24}\end{eqnarray}
where $l=n-{1\over 2}(1+\sigma)$ is an orbital quantum number, and 
\begin{equation}
C_q \equiv \left( {(\varepsilon_q + \varepsilon_q^0)
(\varepsilon_q^0+m) \over 4\varepsilon_q^0 \varepsilon_q} \right)^{1/2},
\end{equation}
\begin{equation}
\rho_\parallel \equiv p_\parallel / (\varepsilon_q + \varepsilon_q^0),
\qquad
\rho_n \equiv p_n / (\varepsilon_q^0 + m),
\end{equation}
\begin{equation}
{\cal A} = \rho_\parallel' + \rho_\parallel,
\qquad
{\cal B} = 1 + \rho_\parallel' \rho_\parallel,
\end{equation}
\begin{equation}
{\cal C} = \rho_\parallel' - \rho_\parallel,
\qquad
{\cal D} = 1 - \rho_\parallel' \rho_\parallel,
\end{equation}
and $p_n = \sqrt{2 n B}$, $\varepsilon_q^0 \equiv \varepsilon(0,n)$,
and the primed versions of the same quantities have unprimed variables
replaced by primed variables. Note again that $\epsilon' p_\parallel' =
\epsilon p_\parallel - k_\parallel$ is implicit throughout.

The functions $J^n_v(k_\perp^2/2B)$ in equations~(\ref{eq:21})
to~(\ref{eq:24}) (the argument is suppressed) are generalized
functions related to the Laguerre polynomials and defined by
\begin{equation}
J^n_v(x) = \left[ {n! \over (n+v)!} \right]^{1/2} \exp(-x/2) x^{v/2}
L^v_n(x) = (-1)^v J^{n+v}_{-v}(x), \label{eq:Jdef}
\end{equation}
where $L^v_n(x)$ are associated Laguerre polynomials.
One may write explicitly
\begin{equation}
J^n_v(x) = \exp(-x/2) x^{v/2} \sum_{m=0}^n (-1)^m {\left[n!(n+v)!\right]^{1/2} 
\over (v+m)!(n-m)!m!} x^m. \label{eq:25}
\end{equation}
Note that $J^n_v(x)=0$ if $n<0$ or $v<-n$. A summary of the properties
of the $J$ functions may be found in \cite{melrose}.

\subsection{Annihilation rate}
We now use the imaginary parts of equations (\ref{eq:14}) to
(\ref{eq:16}) to determine the rate at which a neutrino annihilates on
an antineutrino in a strongly magnetized electron-positron plasma. The
imaginary parts of these integrals come from the poles in the
denominators of the integrands, where $\omega - \epsilon \varepsilon_q
+ \epsilon' \varepsilon_{q'} = 0$. The contribution of these poles may
be calculated through the Plemelj formula
\begin{equation}
{1 \over \omega - \omega_0} = {\cal P} {1 \over \omega - \omega_0} -
i \pi \delta(\omega - \omega_0),
\end{equation}
where ${\cal P}$ denotes the principle value. The integral over the
delta-function introduced into equations (\ref{eq:14}) to
(\ref{eq:16}) by the Plemelj formula may be done explicitly by
determining the positions of the poles of the integrand. For $\omega >
0$, which is the case for neutrino-antineutrino annihilation, the
denominators of equations (\ref{eq:14}) to (\ref{eq:16}) only have
zeros if $\epsilon=1$, $\epsilon'=-1$, (pair creation), $\epsilon=-1$,
$\epsilon'=-1$ (neutrino pair absorption on a positron), and
$\epsilon=1$, $\epsilon'=1$ (neutrino pair absorption on an
electron). Each of these processes occurs in a different region of the
phase space, determined by the existence of the roots of the energy
conservation equation $\omega - \epsilon \varepsilon_q + \epsilon'
\varepsilon_{q'} = 0$. Writing $\delta = \omega^2 - k_\parallel^2$,
this equation has general solutions of the form
\begin{equation}
p_\parallel = \pm {k_\parallel \over 2} \xi \pm {\omega \over 2}
\Delta^{1/2} \label{roots}
\end{equation}
with
\begin{equation}
\xi = 1 + {2(n-n')B \over \delta},
\end{equation}
and
\begin{equation}
\Delta = \xi^2 - {4(1+2 n B) \over \delta}.
\end{equation}
The roots given by equation (\ref{roots}) are real if $\Delta > 0$,
which is true if and only if
\begin{equation}
0< \delta < \left[\sqrt{1+2n' B}-\sqrt{1+2nB}\right]^2,\label{eq:12}
\end{equation}
which corresponds to a pair absorption on an electron or positron, and
\begin{equation}
\delta > \left[\sqrt{1+2n' B}+\sqrt{1+2nB}\right]^2,\label{eq:27}
\end{equation}
which corresponds to an electron positron pair creation. Note that for
two annihilating neutrinos one has $\delta = \omega^2 - k_\parallel^2 > 0$.

The contributions of the vector-vector, axial-vector and axial-axial
parts of the rates may be separated out through
\begin{equation}
\frac{dR_{\nu \bar \nu}}{d^3q_1d^3q_2}  = {1 \over 2 q_1 2 q_2}
\frac{1}{1-e^{-E/T_e}} M_{\mu\nu} {\rm Im}\{ c_V^2 \alpha^{\mu\nu}(K)
+ 2 c_V c_A \,\alpha^{\mu\nu}_5(K) 
+c_A^2 \,\alpha^{\mu\nu}_{55}(K) \} \label{eq:51}
\end{equation}
with $K = Q_1+Q_2$. Each of the polarization tensor components
in equation (\ref{eq:51}) contains a sum over two sets of Landau
levels evaluated at the root of the denominator, and a sum over the
$\epsilon$ and $\epsilon'$, which denote the type of particles
which are interacting. Here, we divide the rate into the three
contributing combinations of $\epsilon$ and $\epsilon'$,
corresponding to three different physical processes.

\subsubsection{Neutrino pair creation}
If equation~(\ref{eq:27}) holds, and $\epsilon=1$, $\epsilon'=-1$, the
poles in the integrand of the polarization tensor components
correspond to the physical process of neutrino-antineutrino
annihilation to form an electron-positron pair,
$\nu\bar{\nu}\leftrightarrow e^+ e^-$.  Letting $X$ stand for a blank,
5, or 55, the imaginary part of the polarization tensor components may be written
\begin{eqnarray}
{\rm Im}\{\alpha^{\mu\nu}_X(K)\} & =&  {G_F^2 eB \over 4 \pi}
 \sum_{n,n'=0}^\infty  {\varepsilon_q \varepsilon_{q'} \over |
 p_\parallel \varepsilon_{q'} - p_\parallel' \varepsilon_q | }\times
 \nonumber \\
& & \left. \,_XT^{\mu\nu}_{-+} (-1 + n_{e^-}(p_\parallel,n) +
 n_{e^+}(p_\parallel',n')) \right|_{p_\parallel:\omega-\varepsilon_q -
 \varepsilon_{q'}=0}\label{eq:39}
\end{eqnarray}
with $p_\parallel' = -p_\parallel + k_\parallel$. This equation is
combined with equation~(\ref{eq:51}) and equation (\ref{totrate}) to
obtain the net rate of annihilation to neutrino antineutrino pairs to
electron positron pairs. The sum in equation (\ref{eq:39}) is evaluated
at the parallel momenta for which the resonance condition is satisfied
\begin{equation}
p_\parallel =  {k_\parallel \over 2} \xi \pm {\omega \over 2} \Delta^{1/2},
\end{equation}
\begin{equation}
p_\parallel' =  {k_\parallel \over 2} \xi \mp {\omega \over 2} \Delta^{1/2}.
\end{equation}
Equation~(\ref{eq:39}) should be evaluated at both of
these roots, and the sum over the Landau orbitals is limited to $n$
and $n'$ such that $\sqrt{1+2 n B} + \sqrt{1+2 n' B} < \sqrt{\omega^2
-k_\parallel^2}$.

The form taken by the blocking factors in equation (\ref{eq:39}) is
due to the fact that the imaginary part of the polarization tensor is
related to the difference of the forward and reverse rates \cite{weldon1}.

\subsubsection{Gyromagnetic absorption on positrons}
If condition (\ref{eq:12}) holds and if $n'>n$, with $\epsilon=-1$,
and $\epsilon'=-1$, the
neutrino-antineutrino pair may be absorbed on a positron,
$\nu\bar{\nu} e^+\leftrightarrow e^+$, a process which is forbidden in
the absence of a strong magnetic field. The contribution to the
imaginary part of the polarization tensor due to the root corresponding to
this process may be written
\begin{eqnarray}
{\rm Im}\{\alpha^{\mu\nu}_X(K)\} & =&  {G_F^2 e B \over 4 \pi}
 \sum_{n,n'=0}^\infty  {\varepsilon_q \varepsilon_{q'} \over |
 p_\parallel \varepsilon_{q'} - p_\parallel' \varepsilon_q | }\times
 \nonumber \\
& & \left. \,_XT^{\mu\nu}_{--} (- n_{e^+}(p_\parallel,n) +
 n_{e^+}(p_\parallel',n')) \right|_{p_\parallel:\omega+\varepsilon_q -
 \varepsilon_{q'}=0} \label{eq:40}
\end{eqnarray}
with $p_\parallel' = p_\parallel + k_\parallel$. The sum is evaluated
for this process at parallel momenta
\begin{equation}
p_\parallel = - {k_\parallel \over 2}\xi \pm  {\omega \over 2}\Delta^{1/2},
\end{equation}
\begin{equation}
p_\parallel' =  {k_\parallel \over 2} \xi \pm {\omega \over 2} \Delta^{1/2}.
\end{equation}

\subsubsection{Gyromagnetic absorption on electrons}
On the other hand, if condition (\ref{eq:12}) holds and if $n>n'$, and
$\epsilon=1$ and $\epsilon'=1$, the
neutrino-antineutrino pair may be absorbed on an electron,
$\nu\bar{\nu} e^-\leftrightarrow e^-$, a process which is also forbidden in
the absence of a strong magnetic field. The contribution to the
imaginary part of the polarization tensor due to the root corresponding to
this process may be written
\begin{eqnarray}
{\rm Im}\{\alpha^{\mu\nu}_X(K)\} & =&  {G_F^2 e B \over 4 \pi}
 \sum_{n,n'=0}^\infty  {\varepsilon_q \varepsilon_{q'} \over |
 p_\parallel \varepsilon_{q'} - p_\parallel' \varepsilon_q | }\times
 \nonumber \\
& & \left. \,_XT^{\mu\nu}_{++} (n_{e^-}(p_\parallel,n) -
 n_{e^-}(p_\parallel',n')) \right|_{p_\parallel:\omega-\varepsilon_q +
 \varepsilon_{q'}=0} \label{eq:41}
\end{eqnarray}
with $p_\parallel' = p_\parallel - k_\parallel$. The sum in
equation~(\ref{eq:41}) should be evaluated at the roots
\begin{equation}
p_\parallel =  {k_\parallel \over 2} \xi \pm  {\omega \over 2} \Delta^{1/2},
\end{equation}
\begin{equation}
p_\parallel' =  -{k_\parallel \over 2} \xi\pm {\omega \over 2} \Delta^{1/2}.
\end{equation}

For the gyromagnetic absorption processes, the limits on the sums over
the Landau levels enumerated by $n$ and $n'$ are given by
$\left|\sqrt{1+2 n B} - \sqrt{1+2 n' B}\right| > \sqrt{\omega^2
-k_\parallel^2}$. There is no upper limit on either $n$ or $n'$
implied by this relation. Instead, the upper limit on the Landau level
is constrained by the fact that one of the distribution functions in
equations (\ref{eq:40}) and (\ref{eq:41}) must be non-zero for there
to be a rate, and these distribution functions are sampled at the
resonant momenta given above, which grow with increasing Landau
number.

\subsection{Scattering rate}

A similar procedure may be used to calculate the rate of scattering of
a neutrino or antineutrino off a thermal pair plasma in a strong
magnetic field. In this case, the frequency of the disturbance in the
field is given by $\omega = q_1 - q_2$, which may be negative or
positive. Hence, there are four contributions to the scattering rate,
whereas there are only three contributions to the annihilation rate.
The processes correspond to $\epsilon=1$,$\epsilon'=-1$, pair creation
by a neutrino, $\epsilon=1$,$\epsilon'=-1$ pair absorption by a neutrino,
$\epsilon=-1$,$\epsilon'=-1$, neutrino positron scattering, and
$\epsilon=1$,$\epsilon'=1$ neutrino electron scattering.
The rates of these processes are given by equation (\ref{scattotrate}) with
\begin{equation}
\frac{dR_{\nu e^\pm}}{d^3q_1d^3q_2} 
= {1 \over 2 q_1 2 q_2}
\frac{1}{1-e^{-E/T_e}}
M_{\mu\nu} {\rm Im}\{ c_V^2 \alpha^{\mu\nu}(K)
+ 2 c_V c_A \,\alpha^{\mu\nu}_5(K) 
+c_A^2 \,\alpha^{\mu\nu}_{55}(K) \} \label{eq:26}
\end{equation}
and with $K = Q_1-Q_2$. Again, the scattering and absorption processes occur in different
regions of the phase space, with the conditions
\begin{equation}
 \delta = \omega^2 - k_\parallel^2  < \left[\sqrt{1+2n' B}-\sqrt{1+2nB}\right]^2,\label{eq:44}
\end{equation}
which corresponds to neutrino scattering off electrons and positrons, and
\begin{equation}
 \delta  > \left[\sqrt{1+2n' B}+\sqrt{1+2nB}\right]^2,\label{eq:45}
\end{equation}
which corresponds to electron positron pair creation and absorption on
neutrinos. 

\subsubsection{Electron-positron pair creation and absorption}
The contribution of the pair creation and absorption processes to the
total neutrino scattering rate, $\nu e^+ e^- \leftrightarrow \nu$, appears in two
parts, for opposite signs of the energy transfer, $\omega$. For $\omega
= q_1 - q_2 > 0$, one has a contribution for $\epsilon = 1$,
$\epsilon' = -1$ if $\delta > \left[\sqrt{1+2nB} +
\sqrt{1+2n'B}\right]^2$. This contribution is given by
\begin{eqnarray}
{\rm Im}\{\alpha^{\mu\nu}_X(K)\} & =&  {G_F^2 e B \over 4 \pi}
 \sum_{n,n'=0}^\infty  {\varepsilon_q \varepsilon_{q'} \over |
 p_\parallel \varepsilon_{q'} - p_\parallel' \varepsilon_q | }\times
 \nonumber \\
& & \left. \,_XT^{\mu\nu}_{+-} (-1 + n_{e^-}(p_\parallel,n) +
 n_{e^+}(p_\parallel',n')) \right|_{p_\parallel:\omega-\varepsilon_q -
 \varepsilon_{q'}=0} \label{eq:46}
\end{eqnarray}
with $p_\parallel' = -p_\parallel +
k_\parallel$. Equation~(\ref{eq:46}) should be evaluated at both the
roots given by,
\begin{equation}
p_\parallel =  {k_\parallel \over 2} \xi \pm  {\omega \over 2} \Delta^{1/2},
\end{equation}
\begin{equation}
p_\parallel' =  {k_\parallel \over 2} \xi \mp {\omega \over 2} \Delta^{1/2}.
\end{equation}

For negative energy transfer, $\omega = q_1 - q_2 < 0$, one has a contribution for $\epsilon = -1$,
$\epsilon' = 1$ if $\delta > \left[\sqrt{1+2nB} +
\sqrt{1+2n'B}\right]^2$. This contribution is given by
\begin{eqnarray}
{\rm Im}\{\alpha^{\mu\nu}_X(K)\} & =&  {G_F^2 e B \over 4 \pi}
 \sum_{n,n'=0}^\infty  {\varepsilon_q \varepsilon_{q'} \over |
 p_\parallel \varepsilon_{q'} - p_\parallel' \varepsilon_q | }\times
 \nonumber \\
& & \left. \,_XT^{\mu\nu}_{-+} (1 - n_{e^+}(p_\parallel,n) -
 n_{e^-}(p_\parallel',n')) \right|_{p_\parallel:\omega+\varepsilon_q +
 \varepsilon_{q'}=0} \label{eq:47}
\end{eqnarray}
with $p_\parallel' = -p_\parallel - k_\parallel$.
Equation~(\ref{eq:47}) should be evaluated at both the
roots given by,
\begin{equation}
p_\parallel =  -{k_\parallel \over 2} \xi \pm  {\omega \over 2} \Delta^{1/2},
\end{equation}
\begin{equation}
p_\parallel' =  -{k_\parallel \over 2} \xi \mp {\omega \over 2} \Delta^{1/2}.
\end{equation}

\subsubsection{Neutrino-electron scattering}
There is a contribution to the imaginary part of the polarization tensor
for $\epsilon = \epsilon' = 1$, if $\delta <
\left[\sqrt{1+2nB} + \sqrt{1+2 n'B}\right]^2$. That is, if this
condition holds, the neutrino-electron scattering is kinematically
allowed, and it contributes to the total neutrino scattering rate through
\begin{eqnarray}
{\rm Im}\{\alpha^{\mu\nu}_X(K)\} & =&  {G_F^2 e B \over 4 \pi}
 \sum_{n,n'=0}^\infty  {\varepsilon_q \varepsilon_{q'} \over |
 p_\parallel \varepsilon_{q'} - p_\parallel' \varepsilon_q | }\times
 \nonumber \\
& & \left. \,_XT^{\mu\nu}_{++} (n_{e^-}(p_\parallel,n) -
 n_{e^-}(p_\parallel',n')) \right|_{p_\parallel:\omega-\varepsilon_q +
 \varepsilon_{q'}=0} \label{eq:49}
\end{eqnarray}
with $p_\parallel' = p_\parallel - k_\parallel$.
If $\delta <0$, then there is only one root to the
equation $\omega-\varepsilon_q + \varepsilon_{q'}=0$, which lies at
\begin{equation}
p_\parallel =  {k_\parallel \over 2}\xi +  {\omega \over 2}\Delta^{1/2},\label{eq:52}
\end{equation}
if $k_\parallel > 0$, and at
\begin{equation}
p_\parallel =  {k_\parallel \over 2}\xi -  {\omega \over 2}\Delta^{1/2},\label{eq:53}
\end{equation}
if $k_\parallel < 0$.  If, on the other hand, $0 < \delta < \left[\sqrt{1+2nB} + \sqrt{1+2n'B}\right]^2$, then
there are possibly two roots or no roots, depending on the relative
sign of $\omega+k_\parallel$ and $n'-n$. If $n'-n$ has the same sign as
$\omega+k_\parallel$, then there are no roots, and this scattering
reaction is not kinematically allowed. If they are different in sign,
then both the solutions given by equations~(\ref{eq:52}) and
(\ref{eq:53}) are roots to the equation and represent open scattering
channels.

\subsubsection{Neutrino-positron scattering}
The structure of the theory for neutrino-positron scattering is
similar to that of neutrino-electron scattering. The rate of
neutrino positron scattering may be written, 
\begin{eqnarray}
{\rm Im}\{\alpha^{\mu\nu}_X(K)\} & =&  {G_F^2 e B \over 4 \pi}
 \sum_{n,n'=0}^\infty  {\varepsilon_q \varepsilon_{q'} \over |
 p_\parallel \varepsilon_{q'} - p_\parallel' \varepsilon_q | }\times
 \nonumber \\
& & \left. \,_XT^{\mu\nu}_{--} (- n_{e^+}(p_\parallel,n) +
 n_{e^+}(p_\parallel',n')) \right|_{p_\parallel:\omega+\varepsilon_q -
 \varepsilon_{q'}=0}  \label{eq:48}
\end{eqnarray}
with $p_\parallel' = p_\parallel + k_\parallel$. If $\omega^2 -
k_\parallel^2 < 0$ then the energy conservation equation, $\omega+\varepsilon_q -
 \varepsilon_{q'}=0$ has a single root given by
\begin{equation}
p_\parallel =  -{k_\parallel \over 2}\xi +  {\omega \over 2} \Delta^{1/2},\label{eq:54}
\end{equation}
if $k_\parallel > 0$, and at
\begin{equation}
p_\parallel =  -{k_\parallel \over 2} \xi -  {\omega \over 2} \Delta^{1/2},\label{eq:55}
\end{equation}
if $k_\parallel < 0$.

Alternately, if $0 < \delta < \left[\sqrt{1+2nB} +
\sqrt{1+2n'B}\right]^2$, then there are possibly two roots or no
roots, depending on the relative sign of $\omega+k_\parallel$ and
$n'-n$. In this case, if $n'-n$ has the opposite sign to
$\omega+k_\parallel$, then there are no roots, and this scattering
reaction is not kinematically allowed. If they have the same sign, then
both the solutions given by equations~(\ref{eq:54}) and (\ref{eq:55})
are roots to the equation and represent open neutrino-positron
scattering channels.

\section{Explicit evaluation of the rates}
In equations~(\ref{eq:39}) to (\ref{eq:41}) and
equations~(\ref{eq:46}) to (\ref{eq:48}) we have expressions for the
rates of the various neutrino-electron processes in terms of double
sums over the electronic Landau orbitals. In general, these
expressions may only be evaluated numerically. To demonstrate how this
is done, we now make a simple calculation of the rate of
electron-positron pair production through neutrino-antineutrino pair
annihilation in such a strong field that only electrons and
positrons in the lowest Landau orbital may be produced. This is not a
physically interesting calculation, as the approximations made to
allow an analytic calculation are quite extreme.

\subsection{Emission into the lowest Landau orbital}
To illustrate the use of equation~(\ref{eq:51}) we calculate the rate
at which electrons and positrons are created in the lowest Landau
orbital by neutrino-antineutrino annihilation for head-on
collisions. 

We assume that the magnetic field lies in the $z$-direction, and
let the neutrino 4-momenta be given by
$Q_1=(q_1,q_1\sin\alpha,0,q_1\cos\alpha)$ which lies in the $x$-$z$ plane,
and
$Q_2=(q_2,q_2\sin\theta\cos\phi,q_2\sin\theta\sin\phi,q_2\cos\theta)$.
As the electron and positron are created in the lowest Landau orbital,
we have that $n=n'=0$, $l=l'=0$, and thus, $\sigma=\sigma'=-1$. This
then leads to the following expressions for the vertex functions,
\begin{equation}
\Gamma^0 = -C_q C_q'(\rho_\parallel'+\rho_\parallel) J_0^0,\label{eq:28}
\end{equation}
\begin{equation}
\Gamma^3 = -C_q C_q'(1 + \rho_\parallel'\rho_\parallel) J_0^0,
\end{equation}
\begin{equation}
\,_5\Gamma^0 = C_q C_q'(1+\rho_\parallel'\rho_\parallel) J_0^0,
\end{equation}
\begin{equation}
\,_5\Gamma^3 = C_q C_q'(\rho_\parallel'+\rho_\parallel) J_0^0\label{eq:29}
\end{equation}
with the other components zero.

Using equations~(\ref{eq:28}) to (\ref{eq:29}), the contraction over
the neutrino tensor in equation (\ref{eq:51}) may be performed
explicitly, leading to the rate of production of electron positron
pairs in the lowest Landau orbital from a neutrino-antineutrino pair,
\begin{eqnarray}
{d R_{\nu \bar \nu} \over d^3\,q_1\  d^3\, q_2} &=& { G_F^2 eB \over
2 \pi}  C_q^2 C_q^{'2} E^{-k_\perp^2/2B} {\varepsilon_q \varepsilon_{q'} \over |
 p_\parallel \varepsilon_{q'} - p_\parallel' \varepsilon_q | } \nonumber \\
& & \times [(1 + \cos\theta\cos\alpha
+ \sin\theta\sin\alpha\cos\phi)(c_V A - c_A B)^2  \nonumber \\ & & +
2(\cos\theta+\cos\alpha)(c_V A-c_A B)(c_A A -c_V B)
\nonumber \\ & & +(1 +  \cos\theta\cos\alpha- \sin\theta\sin\alpha\cos\phi)
(c_V B - c_A A)^2], \label{eq:30}
\end{eqnarray}
where $A = \rho_\parallel + \rho_\parallel'$, $B = 1 +
\rho_\parallel\rho_\parallel'$, and $(J^0_0)^2=e^{-k_\perp^2/2B}$ has
been used.

Equation (\ref{eq:30}) is to be evaluated at both of the solutions to $\omega =
\varepsilon(p_\parallel,0) +
\varepsilon(-p_\parallel+k_\parallel,0)$. These are given by
\begin{equation}
p_\parallel = {k_\parallel \over 2} \pm {\omega \over 2} \sqrt{1 - {4
\over \omega^2 - k_\parallel^2}}
\end{equation}
with
\begin{equation}
\omega = q_1+q_2, \qquad k_\parallel^2 = q_1^2\cos^2\alpha + q_2^2
\cos^2\theta + 2 q_1 q_2 \cos\alpha \cos\theta.
\end{equation}
Also,
\begin{equation}
p_\parallel' = {k_\parallel \over 2} \mp {\omega \over 2} \sqrt{1 - {4
\over \omega^2 - k_\parallel^2}}.
\end{equation}

We further restrict our attention to head-on collisions between a neutrino and
antineutrino with the same energy, $q_1=q_2=q$. Leaving the angle between the neutrino and
the magnetic field as $\alpha$, the antineutrino angles must be
$\theta=\pi-\alpha$, and $\phi=\pi$. Under these conditions,
$k_\parallel=0$, and $k_\perp=0$, as we are in the centre of momentum
frame. Thus, $p_\parallel=\pm\sqrt{q^2-1}$, and
$p_\parallel'=-p_\parallel$. Evaluating equation~(\ref{eq:30})
leads to
\begin{equation}
{d R_{\nu \bar \nu} \over d^3q_1  d^3q_2} = {G_F^2 c_V^2 e B \over
\pi} {m_e^2 \over q \sqrt{q^2 - 1}} \sin^2\alpha,\label{eq:31}
\end{equation}
where there is a factor of 2 due to the two roots of the resonant
denominator. Note the dependence on the angle with respect to the
magnetic field. This shows the intrinsic anisotropy of this process, in
that neutrinos approaching head-on along the field line cannot
interact to produce pairs.

\subsection{General Calculation}
In general, the calculation of the rates proceeds numerically. Our
simple calculation motivates the numerical approach. One specifies the
incident neutrino energies and angles, and explicitly calculates the
leptonic tensor, equation (\ref{leptens}). Then, for each allowed $n$
and $n'$ for the process in question, one calculates the location of
the roots, and the vertex functions, equations~(\ref{eq:21}) to
(\ref{eq:24}), evaluated at these roots. From the vertex functions,
one constructs the $\,_X T^{\mu\nu}_{\epsilon\epsilon'}$ matrices and
contracts them with the leptonic tensor. The results of the sum are
then premultiplied with the phase space factors and other prefactors
to give the differential rate of the process.

While this is a reasonably complicated procedure, it has a certain
natural advantage, as it exploits the similarities between the
calculations of the different processes. The alternate method of
Bezchastnov and Haensel \cite{bezch} is not essentially simpler. The procedure
outlined here can be performed in such a way that the rates for all
seven processes can be calculated simultaneously, with the
common elements of the calculation shared for computational
efficiency.

\subsection{Weak magnetic field limit}

\begin{figure}
\centerline{\epsfxsize=8cm\epsfbox{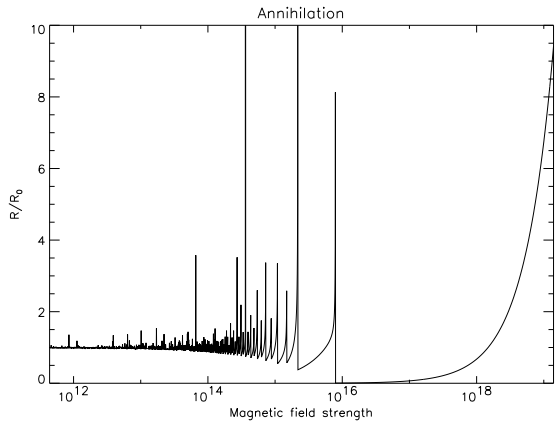}}
\caption{Rate of neutrino-antineutrino annihilation for a nearly
head-on collision between a 5 MeV neutrino and a 6 MeV antineutrino in
vacuum as a function of magnetic field. The rate has been scaled
against the unmagnetized rate, and shows that the unmagnetized theory
is reproduced in the limit of a small magnetic field. The behaviour of
the rate for very high field strengths where only particles in the
lowest Landau orbitals are produced is linear, and follows the results
derived in section 4A.}
\end{figure}

Thus far, it has proved impossible to show analytically that the
expressions given above reduce to the unmagnetized rate in the limit
of a weak magnetic field. However, the numerical calculation of the
rates does reproduce the expected behaviour in the weak field
limit. In figure 3, we show the rate of electron-positron pair
production through neutrino annihilation as a function of magnetic
field strength, normalized to the unmagnetized rate. This plot
shows that the unmagnetized limit is obtained in the limit of small
magnetic field strength, as the graph tends to unity for a small
magnetic field.

\subsection{Numerical results}
We turn now to a more general survey of the properties of the
annihilation rate as a function of magnetic field strength, neutrino
energy, plasma temperature, and the various angles in the
system. Throughout we assume that the chemical potential of the plasma
is zero. In general, the rates are very complicated functions of these
variables. To determine the effect of a strong magnetic field on an
astrophysical system, these rates must be averaged over thermal
distributions of neutrinos and over a realistic angular
distribution. This exercise is beyond the scope of this work.

In figure 4 we show the rate of neutrino-antineutrino annihilation,
scaled to the unmagnetized rate, of a 4 MeV neutrino and a 5 MeV
antineutrino in a pair plasma of temperature $T=2$ MeV as a function of magnetic field
strength. This plot shows the fine structure that comes about due to
the quantized perpendicular momentum states. The underlying rate of
electron-positron pair production is a weak function of the field
strength until around $10^{15}$ G. This is the field strength at which
$B\approx q_1 + q_2$ in scaled coordinates, and thus, the magnetic
field begins to dominate above this level. It is also here that the
rate of gyromagnetic absorption of neutrino-antineutrino pairs becomes
significant - this is because the amount of energy required to lift an
electron or positron from its lowest Landau level to another (low)
Landau level is more closely matched to the available neutrino
energy. This allows the phase space for this reaction to grow and for
the process to proceed more vigorously. Consequently, the effect of
the magnetic field is stronger for low energy neutrinos.

In figure 5 we show the annihilation rate for the same conditions as
figure 4, except that the magnetic field is fixed at $4 \times
10^{15}$ G and the energy of the annihilating neutrino is allowed to
vary. This plot shows the standard increase in cross section as a
function of energy for the pair creation rate, but also shows a number
of other interesting features. For instance, for low energy neutrinos,
the pair creation rate is suppressed - this is because there is only
enough energy to create pairs in the lowest Landau level, and these
states are filled by the thermal plasma. On the other hand, there is a
significant rate of gyromagnetic absorption on the plasma particles in
the lowest Landau orbital, even for annihilation with the lowest
energy neutrinos. Of course, these rates must be averaged over thermal
distributions of neutrinos and antineutrinos, so it is very unclear as
to which of the processes will dominate for which regions of parameter
space.

Figures 6 to 8 show the annihilation rates as a function of the
angles of the system. They show that the annihilation rates are
anisotropic, and can have a very sensitive dependence on angle. At
higher neutrino energies or at lower field strengths there can be many
more resonances, leading to highly anisotropic heating and momentum
deposition.

\begin{figure}
\centerline{\epsfxsize=8cm\epsfbox{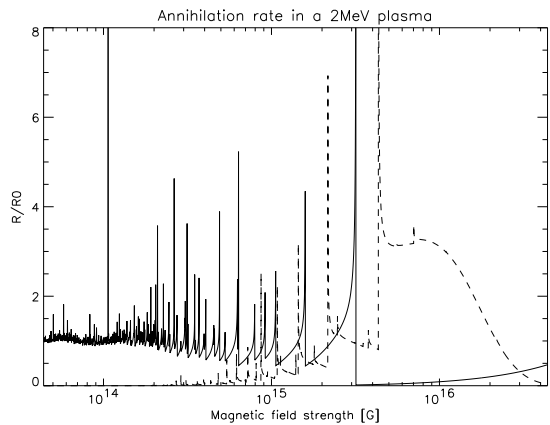}}
\caption{Representative calculation calculation of the rate of
annihilation of a 4 MeV neutrino and an 5 MeV antineutrino in a 2 MeV
pair plasma as a function of magnetic field strength. The solid line
shows the rate at which the pair process occurs as a function of field
strength, and the dashed line shows the rate of gyromagnetic
absorption onto electrons and positrons. The angular variables were
chosen such that $\alpha = 0.2$, $\theta=1.1$, and $\phi = 0.4$.}
\end{figure}

\begin{figure}
\centerline{\epsfxsize=8cm\epsfbox{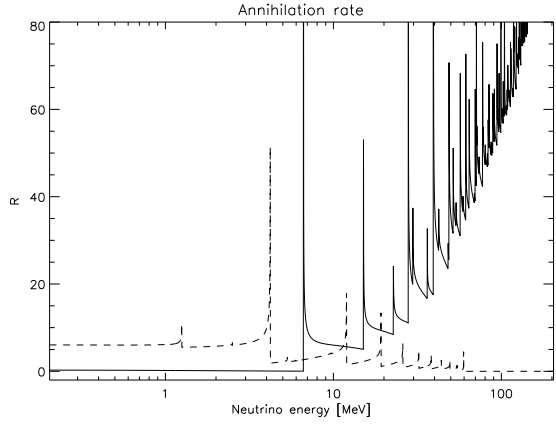}}
\caption{The rate of neutrino-antineutrino annihilation in a 2 MeV
pair plasma as a function of the neutrino energy. The magnetic field
is fixed at $4 \times 10^{15}$ G, and the other variables are as in
figure 4.}
\end{figure}

\begin{figure}
\centerline{\epsfxsize=8cm\epsfbox{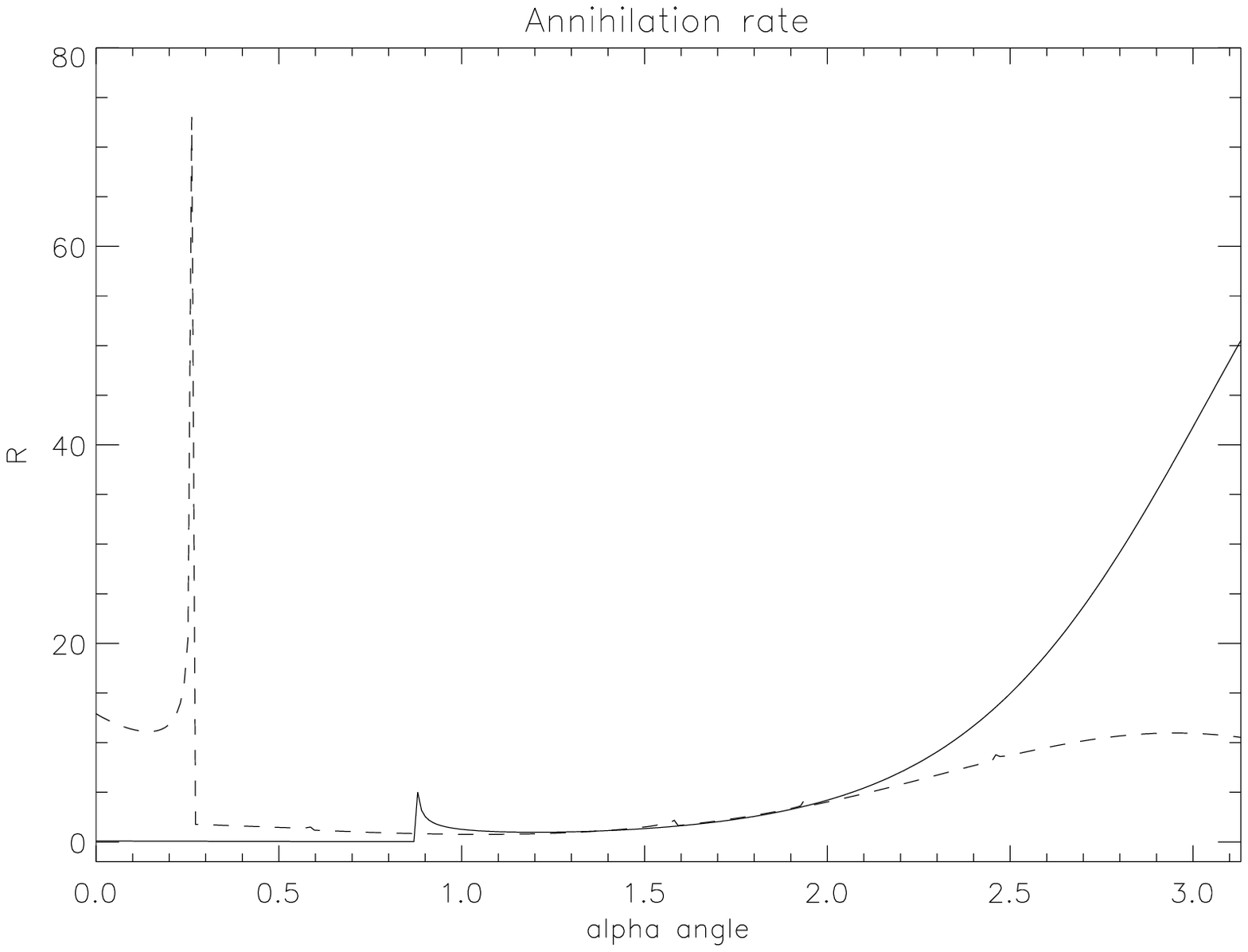}}
\caption{The rate of neutrino-antineutrino annihilation in a 2 MeV
pair plasma as a function of $\alpha$, the angle between the incoming
neutrino and the magnetic field. The other variables are as in
figures 4 and 5.}
\end{figure}

\begin{figure}
\centerline{\epsfxsize=8cm\epsfbox{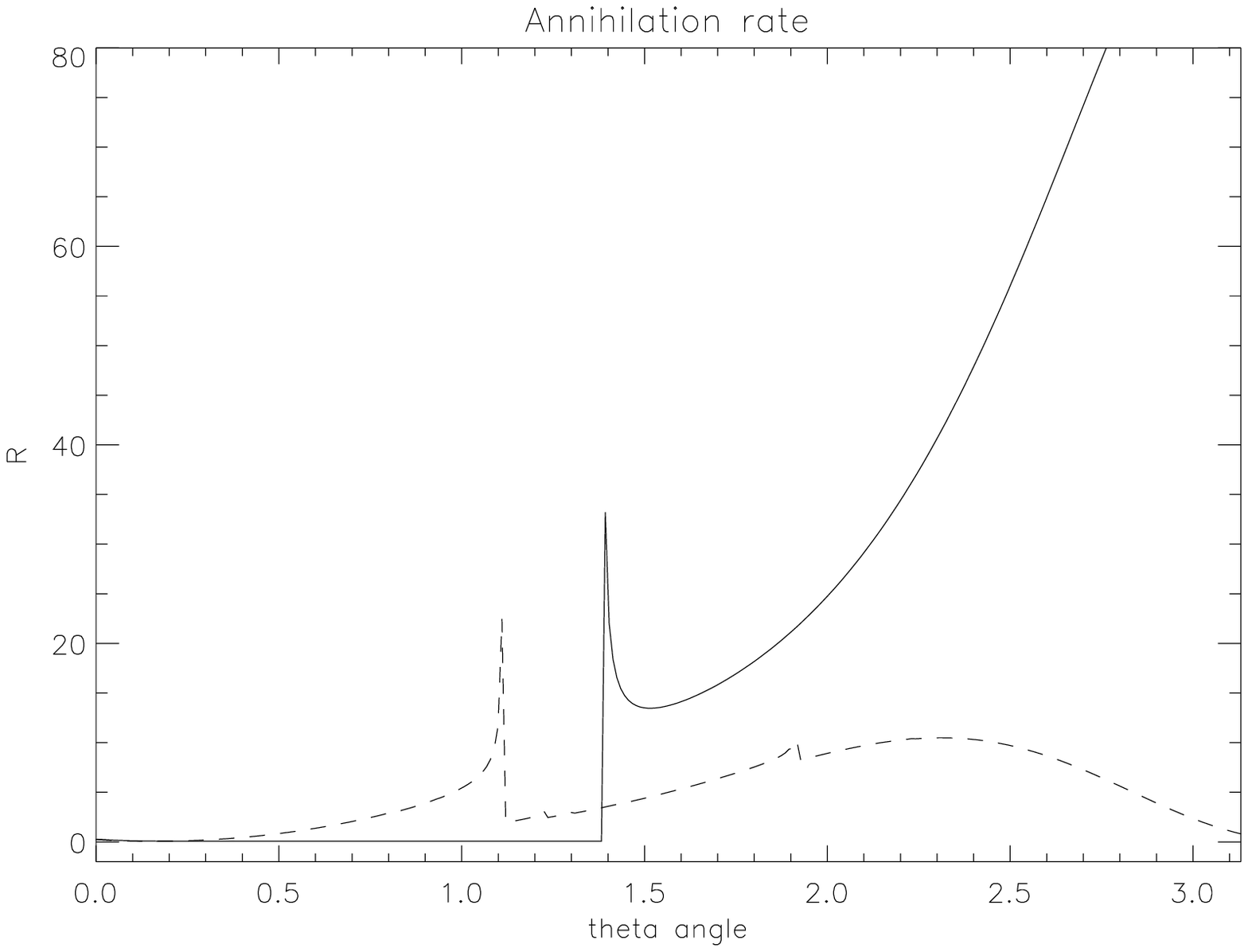}}
\caption{The rate of neutrino-antineutrino annihilation in a 2 MeV
pair plasma as a function of $\theta$, the poloidal angle between the
incoming anti-neutrino and the magnetic field. The other variables are
as in figures 4 and 5.}
\end{figure}

\begin{figure}
\centerline{\epsfxsize=8cm\epsfbox{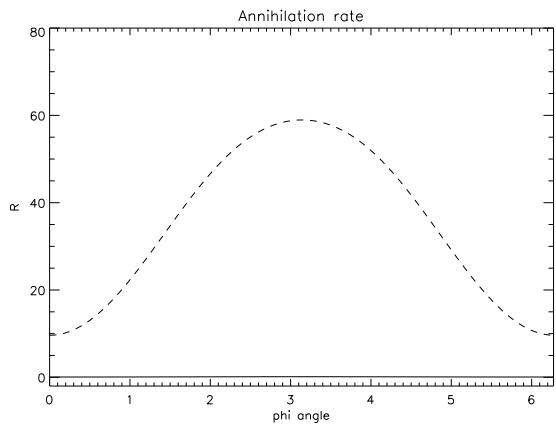}}
\caption{The rate of neutrino-antineutrino annihilation in a 2 MeV
pair plasma as a function of $\phi$, the toroidal angle between the
incoming anti-neutrino and the magnetic field. The other variables are
as in figures 4 and 5.}
\end{figure}

\section{Conclusions}
We have presented here a new calculation of the rates of
neutrino-electron interactions in a strongly magnetized thermal
electron-positron plasma using FTFT. Starting from the imaginary part of the
polarization tensor all neutrino-electron processes, allowed in a strong
magnetic field, can be treated simultaneously. 
This calculation is accurate to first order
in the Fermi theory, treats the effect of the magnetic field on the
wave functions of the electrons exactly, and includes the effects of
Pauli blocking.

These calculations show that the rate of the standard unmagnetized
processes may be modified greatly by the presence of a strong magnetic
field, and that processes which only occur in a strong magnetic field
are the dominant energy deposition processes for a range of field
strengths and neutrino and antineutrino parameters.

To make stronger conclusions about the physical implications of these
processes in strong magnetic fields it is necessary to average the
rates calculated here over thermal distributions of neutrinos
appropriate to the astrophysical scenarios being considered. However,
it is clear that there must be a reconsideration of the role strong
magnetic fields and neutrinos may play in extremely energetic
astrophysical events such as core collapse supernovae and gamma ray
bursts.

\section*{Acknowledgements}
S.J.H. was supported by the TMR and the Deutsche
Forschungsgemeinschaft project SFB 375-95, and by the Australian
Research Council. M.H.T. was supported
by the Deutsche Forschungsgemeinschaft.

\end{document}